\documentclass[12pt]{JHEP3}
\newcommand{\be}[1]{ \begin{equation}\label{#1} }
\newcommand{\ee}{\end{equation}}
\newcommand{\bea}[1]{\begin{eqnarray}\label{#1} }
\newcommand{\eea}{\end{eqnarray}}

\def\ZZZ{{\hskip-3pt\hbox{ Z\kern-1.6mm Z}}}
\def\zzz{{\hskip-3pt\hbox{ z\kern-1mm z}}}
\usepackage{amsmath}
\usepackage{amstext,amssymb,amsfonts}
\newcommand{\gl}{\lambda}

\newcommand{\refb}[1]{(\ref{#1})}

\def\one{{\hbox{ 1\kern-.8mm l}}}
\def\zero{{\hbox{ 0\kern-1.5mm 0}}}

\def\t{\tau}

\def\cj{{\mathcal J}}
\def\ck{{\mathcal K}}

\def\cu{{\mathcal U}}

\def\bz{{\mathbb Z}}
\def\tr{{\rm Tr}}

\def\l{\left}
\def\r{\right}

\def\D{\Delta}
\def\s{\sigma}
\def\g{\gamma}
\def\G{\Gamma}

\def\o[#1]{{\rm O}\left({#1}\right)}
\def\dotl[#1,#2]{\left\langle #1, #2 \right\rangle}
\def\dotlb[#1,#2]{[ #1, #2 ]}
\def\dotp[#1,#2]{(#1) \cdot (#2)}
\def\>{\rangle}
\def\<{\langle}
\def\tl{\tilde{l}}
\def\tmu{\tilde{\mu}}

\title{The Heat Kernel on $AdS$}

\author{Rajesh Gopakumar$^{a}$\footnote{gopakumr AT hri DOT res DOT in}, Rajesh Kumar Gupta$^{a,b}$\footnote{R.K.Gupta AT uu DOT nl} and Shailesh Lal$^{a}$\footnote{shailesh AT hri DOT res DOT in} \\
$^a$Harish-Chandra Research Institute, \\
$\;$Chhatnag Road,\\
$\;$Jhusi, India 211019\\
$^b$Institute of Theoretical Physics,\\
$\;$Utrecht University, Leuvenlaan 4, 3584 CE,\\
$\;$Utrecht, The Netherlands\\}

\abstract{We explicitly evaluate the heat kernel for the Laplacian of arbitrary spin tensor fields on the thermal quotient of (Euclidean) $AdS_N$ for $N\geq 3$ using the group theoretic techniques employed for $AdS_3$ in arXiv:0911.5085. Our approach is general and can be used, in principle, for other quotients as well as other symmetric spaces.}

\preprint{HRI/ST/1104}
\begin{document}

\section{Introduction}

Quantum effects in $AdS$ spacetimes are important in the AdS/CFT correspondence in trying to go beyond the conventional tree level gravity description. The leading effects come at one loop and capture, for instance,  important information about the spectrum of the theory. In the language of the boundary gauge theory these come from diagrams which are suppressed by ${1\over N^2}$ compared to the planar ones. For example, this genus one contribution to the free energy of the QFT captures all the quadratic fluctuations about the bulk $AdS$ background.  From a worldsheet perspective, for the closed string theory in the bulk, this is the torus contribution to the free energy. 

The string theory sigma model with an $AdS$ target space is as yet fairly intractable at the quantum level. Thus some important clues about the worldsheet structure might be gained from knowing the spectrum of the gauge theory and trying to reproduce the one loop answer coming from the quadratic fluctuations of the dual fields. Typically the quadratic terms involve Laplacians acting on arbitrary tensor fields. Evaluation of the path integral at quadratic level then requires one to compute the determinants of such general laplacians about an $AdS$ background (more precisely a thermal quotient to define a generating function). The best way to evaluate these is the heat kernel method. This, in addition, has the virtue that the proper time that enters here has an interpretation in terms of the modulus of the genus one worldsheet \cite{Polchinski:1985zf}. Essentially the heat kernel captures the first quantised description of the particles which go into constituting the string spe
 ctrum. 

Recall that given the normalised eigenfunctions $\psi^{(S)}_{n,a}\l(x\r)$ of the Laplacian $\D_{\l(S\r)}$ for a spin-$S$ field on a manifold ${\mathcal M}_{d+1}$, and the spectrum of eigenvalues $E_n^{(S)}$, we can define the heat kernel between two points $x$ and $y$ as
\begin{equation}
\label{genkernel}
K^{(S)}_{ab}\l(x,y;t\r) = \langle y,b | e^{t\Delta_{\l(S\r)}} | x,a \rangle = \sum_n \psi^{(S)}_{n,a}\l(x\r)\psi^{(S)}_{n,b}\l(y\r)^*e^{tE_n^{(S)}},
\end{equation}
where $a$ and $b$ are local Lorentz indices for the field. We can trace over the spin and spacetime labels to define the traced heat kernel as
\begin{equation}
\label{intkernel}
K^{(S)}\l(t\r)\equiv \tr e^{t\Delta_{\l(S\r)}} =\int_{\mathcal M}\sqrt{g} d^{d+1}x\sum_a K^{(S)}_{aa}\l(x,x;t\r).
\end{equation}
The one-loop partition function is related to the trace of the heat kernel through
\begin{equation}
ln\,Z^{(S)}=\ln\,det\l(-\Delta_{\l(S\r)}\r)= \tr \ln\l(-\Delta_{\l(S\r)}\r) =-\int_{0}^\infty\frac{dt}{t}\tr e^{t\Delta_{\l(S\r)}}.
\end{equation}

For a general manifold, the computation of the heat kernel is a formidable task, even for the scalar Laplacian. Typically one has asymptotic results. However, for symmetric spaces such as spheres and hyperbolic spaces (Euclidean $AdS$) there are many simplifications. This is because these spaces can be realised as cosets $G/H$ and one can therefore use the powerful methods of harmonic analysis on group manifolds. In \cite{David:2009xg} these techniques were used to explicitly 
compute the heat kernel on thermal $AdS_3$ for fields of arbitrary spin. Though \cite{David:2009xg} exploited the group theory, it also used at several places the particular fact that $S^3$ (from which one continued to $AdS_3$) itself is the group manifold $SU(2)$. In fact, many properties of $SU(2)$ were used in intermediate steps and it was not completely obvious how these generalise to higher dimensional spheres or $AdS$ spacetimes. 
 
In this paper, we will generalise the methods of \cite{David:2009xg} to compute the heat kernel for the Laplacian for arbitrary spin tensor fields on 
thermal $AdS$ spacetimes. This therefore includes the cases of $AdS_4$, $AdS_5$ and $AdS_7$ which play a central role in the AdS/CFT correspondence. 
Since we are primarily interested in evaluating the one-loop partition function for a spin-$S$ particle, we shall concentrate on the traced heat kernel though we will see that the techniques are sufficiently general. One may, for instance, also adapt these techniques to evaluate other objects of interest such as the (bulk to bulk) propagator.

We shall 
mostly focus on ${\cal M}=AdS_{2n+1}/\Gamma$ 
where $\Gamma$ is the thermal quotient. 
For practical reasons we will obtain the answer for thermal $AdS$ by analytic continuation of the answer for an appropriate ``thermal quotient" of 
the $N$-sphere.  As explained in \cite{David:2009xg} there are good reasons to believe that this  analytic continuation of the harmonic analysis  works for odd dimensional spheres and hyperboloids (Euclidean AdS). For even dimensional hyperboloids there are some additional discrete representations other than the continuous ones one obtains by a straightforward analytic continuation. However, this additional set of representations does not contribute for Laplacians over a wide class of tensor fields, which in particular include the symmetric transverse traceless (STT) tensors. We provide necessary details in Appendix \ref{evendim}.

We expect that the results here will have many applications.  In the three dimensional case the corresponding results have been used to clarify the quantum nature of topologically massive gravity in $AdS_3$ \cite{Gaberdiel:2010xv}. They were also used to show that the higher spin gauge theories (with spins $s=2, 3 \ldots N$) realise ${\cal W}_N$ symmetry at the quantum level \cite{Gaberdiel:2010ar} - generalising the classical Brown-Henneaux like result for such theories \cite{Campoleoni:2010zq,Henneaux:2010xg}. This was an important input in formulating a duality between these higher spin theories (with additional scalar fields) and ${\cal W}_N$ minimal models \cite{Gaberdiel:2010pz}.   

In the higher dimensional cases of interest here one can apply our results to study the Vasiliev higher spin theories. Such theories are conceivably related to a subsector of free Yang-Mills theories and perhaps the higher spin theories are higgsed in an interesting way in going away from the free theory. 
Evaluating the one loop fluctuations in the bulk can help us in checking these conjectures with more precision. 

The plan of the paper is as follows. In the next section we briefly review the harmonic analysis on homogeneous spaces that is the mainstay of these computations and illustrate it with the 
case of $S^{2n+1}$. In Sec. 3, we describe how to generally consider quotients of these symmetric spaces. Sec. 4 describes the analytic continuation of $S^{2n+1}$ to the Euclidean hyperboloids. We evaluate the coincident heat kernels for the general class of symmetric tensor representations and check against existing results in the literature. In Sec. 5 we obtain the answer for the traced heat kernel on thermal $AdS_{2n+1}$. Sec. 6 contains an application to the scalar one loop partition function. Finally Sec. 7 has some concluding remarks. 


\section{The Heat Kernel on Homogeneous Spaces}

\label{HarmonicAnalysis}

The heat kernel of the spin-$S$ Laplacian\footnote{By spin-$S$ we refer to the representation under which the field transforms under tangent space rotations. In the case of $S^{2n+1}$ or (euclidean) $AdS_{2n+1}$, this will be a representation of $SO(2n+1)$. The Laplacian is that of a tensor field transforming in this representation. In the case of spheres and hyberboloids we will consider the Laplacian with the Christoffel connection.}
 may be evaluated over the spacetime manifold $\mathcal M$ by solving the appropriate heat equation. Alternatively, one may attempt a direct evaluation by constructing the eigenvalues and eigenfunctions of the spin-$S$ Laplacian and carrying out the sum over $n$ that appears in (\ref{genkernel}). Both these methods quickly become forbidding when applied to an arbitrary spin-$S$ field. However, if $\mathcal M$ is a homogeneous space $G/H$, then the use of group-theoretic techniques greatly simplifies the evaluation. The main simplifications arise from the following facts which we will review below and then heavily utilise:
\begin{enumerate}

\item
The eigenvalues $E_n^{(S)}$ of the Laplacian $\Delta_{(S)}$ are determined in terms of the quadratic Casimirs of the symmetry group $G$ and the isotropy subgroup $H$. There is thus a large degeneracy of eigenvalues. 

\item
The eigenfunctions $\psi_{n,a}^{(S)}(x)$ are matrix elements of unitary representation matrices of $G$. 

\item
This enables one to carry out the sum over degenerate eigenstates using the group multiplication properties of the matrix elements. Thus a large part of the sum in (\ref{genkernel}) can be explicitly carried out.

\end{enumerate}
We begin with a brief recollection of some basic facts about harmonic analysis on coset spaces. This will collate the necessary tools with which we evaluate (\ref{genkernel}) and (\ref{intkernel}), and will set up our notation. The interested reader is referred to \cite{Salam:1981xd,Camporesi:1990wm,Camporesi:1995fb} for introduction and details and to \cite{David:2009xg} for explicit examples of these constructions. Given compact Lie groups $G$ and $H$, where $H$ is a subgroup of $G$, the coset space $G/H$ is constructed through the right action of $H$ on elements of $G$
\begin{equation}
G/H=\lbrace gH \rbrace.
\end{equation}
(We will also need to consider left cosets, $\G\backslash G$, where $\G$ will act on elements of $G$ from the left.)

We recall that $G$ is the principal bundle over $G/H$ with fibre isomorphic to $H$. Let $\pi$ be the projection map from $G$ to $G/H$, \textit{i.e.}
\begin{equation}
\pi\l(g\r)=gH\quad\forall g\in G.
\end{equation}
Then a section $\s\l(x\r)$ in the principal bundle is a map 
\begin{equation}
\s: G/H \mapsto G,\quad \hbox{such that } \pi\circ\s=e,
\end{equation}
where $e$ is the identity element in $G$, and $x$ are coordinates in $G/H$. 
A class of sections which will be useful later is of the form
\begin{equation}
\label{thermalex}
\s\l(gH\r)=g_{\circ},
\end{equation} 
where $g_{\circ}$ is an element of the coset $gH$, which is chosen by some well-defined prescription. The so called `thermal section' that we choose in Section \ref{thermalquotient} is precisely of this form.
Let us label representations of $G$ by $R$ and representations of $H$ by $S$. We will sometimes refer to representations $S$ of $H$ as spin-$S$ representations.\footnote{In the case of sphere and hyperboloids, $H$ is isomorphic to the group of tangent space rotations for the manifold $G/H$. See footnote 1.} The vector space which carries the representation $R$ is called $\mathcal V_R$, and has dimension $d_R$, while the corresponding vector space for the representation $S$ is $\mathcal V_S$ of dimension $d_S$.

Eigenfunctions of the spin-$S$ Laplacian are then given by the matrix elements
\begin{equation}
\label{eigenf}
{\psi^{\l(S\r)I}_a}\l(x\r)={{\cu^{\l(R\r)}}\l(\s\l(x\r)^{-1}\r)_a}^I,
\end{equation}
where $S$ is the unitary irreducible representation of $H$ under which our field transforms, and $R$ is any representation of $G$ that contains $S$ when restricted to $H$. $a$ is an index in the subspace $\mathcal V_S$ of $\mathcal V_R$, while $I$ is an index in the full vector space $\mathcal V_R$. Generally, a given representation $S$ can appear more than once in $R$. However, we shall be interested in the coset spaces $SO\l(N+1\r)/SO\l(N\r)$ and $SO\l(N,1\r)/SO\l(N\r)$, for which a representation $S$ appears at most once \cite{Barut:1986dd,Camporesi}. We have therefore dropped a degeneracy factor associated with the index $a$, which appears in the more complete formulae given in \cite{Camporesi}.

The corresponding eigenvalues are given by
\begin{equation}
\label{eigenval}
-E_{R,I}^{(S)}= C_2\l(R\r)- C_2\l(S\r).
\end{equation}
The index $n$ for the eigenvalues of the spin-$S$ Laplacian that appeared in (\ref{genkernel}) is therefore a pair of labels, viz. $\l(R,I\r)$,\footnote{Note that $a$ labels the components of the eigenfunction and is not a part of the index $n$.} where the eigenfunctions that have the same label $R$ but a different $I$ are necessarily degenerate. 
We will therefore drop the subscript $I$ for $E^{(S)}$.

The expression (\ref{genkernel}) for the heat kernel then reduces to
\begin{equation}
\label{kernelintermsofeigenfunctions}
K^{(S)}_{ab}\l(x,y;t\r) = \sum_{R,I}a_R^{(S)}\psi^{(S)}_{(R,I),a}\l(x\r)\psi^{(S)}_{(R,I),b}\l(y\r)^*e^{tE_R^{(S)}},
\end{equation}
where $ a_R^{(S)}=\frac{d_R}{d_S}\frac{1}{V_{G/H}}$ is a normalisation constant (see Appendix \ref{normalisationkernel}).
This can be further simplified by putting in the expression (\ref{eigenf}) for the eigenfunctions.
\begin{eqnarray}
\label{untracedkernel}
K^{(S)}_{ab}\l(x,y;t\r) &=& \sum_{R}\sum_{I=1}^{d_R} a_R^{(S)}{{\cu^{\l(R\r)}}\l(\s\l(x\r)^{-1}\r)_a}^I \l[{{\cu^{\l(R\r)}}\l(\s\l(y\r)^{-1}\r)_b}^I\r]^*e^{tE_R^{(S)}} \nonumber \\
&=& \sum_R  a_R^{(S)} {{\cu^{\l(R\r)}}\l(\s\l(x\r)^{-1}\s\l(y\r)\r)_a}^b e^{tE_R^{(S)}},
\end{eqnarray}
where we have used the fact that the $\mathcal U^{\l(R\r)}$ furnish a unitary representation of $G$. As an aside, we note that this matrix representation of the group composition law is the generalisation of the addition theorem for spherical harmonics on $S^2$ to arbitrary homogeneous vector bundles on coset spaces. 

To establish notation for later use, we define the heat kernel with traced spin indices
\begin{equation}
\label{kernel}
K^{(S)}\l(x,y;t\r)\equiv \sum_{a=1}^{d_S} K^{(S)}_{aa}\l(x,y;t\r)
= \sum_R  a_R^{(S)} \tr_S\l({\cu^{\l(R\r)}}\l(\s\l(x\r)^{-1}\s\l(y\r)\r)\r) e^{tE_R^{(S)}},
\end{equation}
where the symbol 
\begin{equation}
\tr_S\l(\cu\r)\equiv\sum_{a=1}^{d_S}\<a,S\vert\cu\vert a,S\>,
\end{equation}
 and can be thought of as a trace over the subspace $\mathcal V_S$ of $\mathcal V_R$. Note that this restricted trace is invariant under a unitary change of basis of $\mathcal V_S$ and 
 {\it not} invariant under the most general unitary change of basis in $\mathcal V_R$.

\subsection{The Heat Kernel on $S^{2n+1}$}

As a prelude to evaluating the traced heat kernel on the ``thermal quotient" of the odd-dimensional sphere, let us evaluate (\ref{untracedkernel}) for the case without any quotient. 
That is, we focus first on $S^{2n+1}\simeq SO(2n+2)/SO(2n+1)$. We will describe the eigenfunctions (\ref{eigenf}) and define the sum over $R$ explicitly. This will  be useful  when we analytically continue our results to the corresponding hyberbolic space. We begin by recalling some facts from the representation theory of special orthogonal groups.

Unitary irreducible representations of $SO(2n+2)$ are characterised by a highest weight, which can be expressed in the orthogonal basis as the array
\begin{equation}
\label{so2nrep}
R=\l(m_1,m_2,\ldots ,m_n,m_{n+1}\r),\quad m_1\geq m_2\geq \ldots \geq m_n \geq \vert m_{n+1}\vert\geq 0
\end{equation}
where the $m_1\ldots m_{n+1}$ are all (half-)integers.
Similarly, unitary irreducible representations of $SO\l(2n+1\r)$ are characterised by the array
\begin{equation}
S=\l(s_1,s_2,\ldots ,s_n\r),\quad s_1\geq s_2\geq \ldots \geq s_n \geq 0,
\end{equation} where the $s_1\ldots s_{n}$ are all (half-)integers. 

Then the quadratic Casimirs for the unitary irreducible representations for an orthogonal group of
rank $n+1$ can be expressed as (see e.g.\cite{Barut:1986dd,Fulton}).
\begin{equation}
\label{Casimir}
C_2\l(m_1,\ldots,m_{n+1} \r)=m^2+2r\cdot m .
\end{equation}
Here the dot product is the usual euclidean one, and the Weyl vector $r$ is given by 
\begin{equation}
r_i=  \l\{\begin{array}{l} n-i+1\quad \rm{if}~ G=SO\l(2n+2 \r), \\ \l(n+\frac{1}{2}\r)-i\quad \rm{if}~ G=SO\l(2n+1\r),\\ \end{array}\r\}
\end{equation}
where $i$ runs from $1$ to $n+1$. 

Let us now consider the expression (\ref{untracedkernel}) for the spin-$S$ Laplacian on $S^{2n+1}$. The eigenvalues $E_R^{\l(S\r)}$ are given by (\ref{eigenval}) and (\ref{Casimir}) and may be written down compactly as
\begin{equation}
\label{eigenvals2n1}
-E_R^{\l(S\r)}=m^2+2r_{SO\l(2n+2\r)}\cdot m-s^2-2r_{SO\l(2n+1\r)}\cdot s.
\end{equation}
The corresponding eigenfunctions are given by (\ref{eigenf}) where we have to specify which are the representations $R$ of $SO\l(2n+2\r)$ that contain a given representation $S$ of $SO\l(2n+1\r)$. This is determined by the branching rules, which for our case state that a representation $R$ given by (\ref{so2nrep}) contains the representation $S$ if
\begin{equation}
\label{snbranching}
\quad m_1\geq s_1 \geq m_2\geq s_2\geq \ldots \geq m_n\geq s_n\geq\vert m_{n+1}\vert,\quad \l(m_i-s_i\r)\in\bz.
\end{equation}
Using these branching rules, one can show that the expression that appears on the right of (\ref{eigenvals2n1}) is indeed positive definite, so that the eigenvalue itself is negative definite as per our conventions.

These rules further simplify if we restrict ourselves to symmetric transverse traceless (STT) representations of $H$. These tensors of rank $s$ 
correspond to the highest weight $\l(s,0,\ldots,0\r)$. In this case, some of these inequalities get saturated, and one obtains the branching rule
\begin{equation}
\label{snbranchingSTT}
\quad m_1\geq s= m_2\geq 0,
\end{equation}
with all other $m_i, s_i$ zero, \footnote{For the special case of $n=1$, i.e. $S^3$, the branching rule is $m_1\geq s= \vert m_2\vert\geq 0$.} and the equality follows from requiring that $R$ contain $S$ in the maximal possible way. Essentially, this is equivalent to the transversality condition.
The sum over $R$ that appears in (\ref{untracedkernel}) is now a sum over the admissible values of $m_1$ in the above inequality. Thus  if we restrict ourselves to evaluating the heat kernel for STT tensors, then we will be left with a sum over $m_1$ only. 

The expression (\ref{eigenvals2n1}) for the eigenvalue also simplifies in this case. With the benefit of hindsight, we will write this in a form that is suitable for analytic continuation to $AdS$
\begin{equation}
\label{eigenvals2n1stt}
-E_R^{\l(S\r)}=\l(m_1+n\r)^2- s -n^2.
\end{equation}
Using these tools, one can write down a formal expression for the heat kernel for a spin-$S$ particle between an arbitrary pair of points $x$ and $y$ on $S^{2n+1}$ using (\ref{untracedkernel}). This is given by
\begin{equation}
\label{kernelsphere}
K^S_{ab}\l(x,y;t\r) = \sum_{m_i}  \frac{n!}{2\pi^{n+1}}\frac{d_R}{d_S} {{\cu^{\l(R\r)}}\l(\s\l(x\r)^{-1}\s\l(y\r)\r)_a}^b e^{tE_R^{(S)}},
\end{equation}
where we have simply expanded out the sum over $R$ into a sum over the permissible values of $m_i$ determined by (\ref{snbranching}), and inserted the expression for the volume of the $\l(2n+1\r)$-sphere. Expressions for the dimensions $d_R$ and $d_S$ are well known (see for example\cite{Barut:1986dd,Fulton}). We list them here for the reader's convenience.
\begin{equation}
\label{dimensions}
d_R=\prod_{i<j=1}^{n+1}\frac{l_i^2-l_j^2}{\mu_i^2-\mu_j^2},\quad d_S= \prod_{i<j=1}^{n}\frac{\tl_i^2-\tl_j^2}{\tmu_i^2-\tmu_j^2}\prod_{i=1}^n \frac{\tl_i}{\tmu_i},
\end{equation}
where $l_i=m_i+\l(n+1\r)-i$, $\mu_i=\l(n+1\r)-i$, $\tl_i=s_i+n-i+\frac{1}{2}$, and $\tmu_i=n-i+\frac{1}{2}$. As explained above, this expression further simplifies for the STT tensors, and we obtain
\begin{equation}
\label{kernelsttsphere}
K^{S}_{ab}\l(x,y;t\r) = \sum_{m_1} \frac{n!}{2\pi^{n+1}} \frac{d_{\l(m_1,s\r)}}{d_s} {{\cu^{\l(m_1,s\r)}}\l(\s\l(x\r)^{-1}\s\l(y\r)\r)_a}^b e^{tE_R^{(S)}},
\end{equation}
where the labels $\l(m_1,s\r)$ and $s$ that appear on the RHS are shorthand for $R=\l(m_1,s,0\ldots,0\r)$ and  $S=\l(s,0\ldots,0\r)$ respectively. This expression should be compared with the equation $\l(3.9\r)$ obtained in \cite{David:2009xg} for the case of $3$ dimensions. The traced heat kernel for the STT tensors is then given by
\begin{equation}
\label{kerneltracesttsphere}
K^{S}\l(x,y;t\r) = \sum_{m_1} \frac{n!}{2\pi^{n+1}} \frac{d_{\l(m_1,s\r)}}{d_s} \tr_S\l(\cu^{\l(m_1,s\r)}\l(\s\l(x\r)^{-1}\s\l(y\r)\r)\r) e^{tE_R^{(S)}}.
\end{equation}
As mentioned earlier, we can in principle use these formulae to construct explicit expressions for the heat kernel between two points $\grave{a}$ \textit{la} \cite{David:2009xg}, and thus for the bulk to bulk propagator. This would, however also require using explicit matrix elements of $SO\l(2n+2\r)$ representations, and we shall not pursue this direction further here.

\section{The Heat Kernel on Quotients of Symmetric Spaces}

We consider the heat kernel on the quotient spaces $\G\backslash G/H$ where $\G$ is a discrete group which can be embedded in $G$. Though it is not essential to our analysis, we will assume that $\G$ is of finite order and is generated by a single element. This is indeed true for the quotients (\ref{quotient}) we consider here. In particular, for the ``thermal quotient" on the $N$-sphere, $\G$ is isomorphic to ${\mathbb Z}_N$.
To evaluate the heat kernel (\ref{genkernel}) on this space, a choice of section that is \textit{compatible} with the quotienting by $\G$ is useful. By this we mean that if 
$\g\in\G$ acts on points $x=gH \in G/H$ by $\g:gH \mapsto\g\cdot gH$, then a section $\s\l(x\r)$ is said to be compatible with the quotienting $\G$ iff 
\begin{equation}
\label{compatible}
\s\l(\g\l(x\r)\r) = \g\cdot\s\l( x\r). 
\end{equation}
The utility of this choice of section will become clear when we explicitly evaluate the traced heat kernel (\ref{intkernel}) for such geometries.

\subsection{The Thermal Quotient of $S^{5}$}
\label{thermalquotient}
As an example, we consider the thermal quotient of $S^5$. This will serve as a useful prototype to keep in mind. We shall also see that we can extrapolate the analysis to the general odd-dimensional sphere. To begin with, let us express the thermal quotient in terms of ``triple-polar" coordinates on $S^5$, which are complex numbers $\l(z_1,z_2,z_3\r)$ such that
\begin{equation}
|z_1|^2+|z_2|^2+|z_3|^2=1.
\end{equation}

We consider the quotient
\begin{equation}
\label{quotient}
\gamma: \{\phi_i \} \mapsto \{ \phi_i+\alpha_i \}.
\end{equation}
Here $\phi_1,\phi_2,\phi_3$ are the phases of the $z$'s and $n_i\alpha_i=2\pi$ for some $n_i\in\mathbb{Z}$ and not all $n_i$s are simultaneously zero.\footnote{Note that this is a more general identification than the  thermal quotient we will need, where one can take $\alpha_i=0\, (\forall i\neq 1)$} However, to embed $\G$ in $SO\l(6\r)$, it is more useful to decompose these complex numbers into 6 real coordinates that embed the $S^5$ into ${\mathbb R}^6$,
\begin{eqnarray}
\label{s5pt}
x_1=\cos\theta \, \cos\phi_1 &\quad& x_2=\cos\theta \, \sin\phi_1, \nonumber \\
x_3=\sin\theta \, \cos\psi \,  \cos\phi_2 &\quad& x_4=\sin\theta \, \cos\psi \, \sin\phi_2, \nonumber \\
x_5=\sin\theta \, \sin\psi \, \cos\phi_3 &\quad& x_6=\sin\theta \,  \sin\psi \,  \sin\phi_3. 
\end{eqnarray}
Now, we construct a coset representative in $SO\l(6\r)$ for this point $x$ with coordinates as in $(\ref{s5pt})$. To do so, we start with the point $\l(1,0,0,0,0,0\r)$ in $R^6$, and construct a matrix $g\l(x\r)$ that rotates this point, the north pole, to the generic point $x$. By construction, $g\l(x\r) \in SO\l(6\r)$, and there is a one-to-one correspondence between the points $x$ on $S^5$ and matrices $g\l(x\r)$, upto a right multiplication by an element of the $SO\l(5\r)$ which leaves the north pole invariant. Such a representative matrix $g\l(x\r)$ can be taken to be
\begin{equation}
\label{cosetrep}
g\l(x\r)=e^{i\phi_1Q_{12}}e^{i\phi_2Q_{34}}e^{i\phi_3Q_{56}}e^{i\psi Q_{35}}e^{i\theta Q_{13}},
\end{equation}
where $Q$'s are the generators of $SO\l(6\r)$. This is clearly an instance of a section in $G$ over $G/H$.

The action of the thermal quotient (\ref{quotient}) on the coset representative $g\l(x\r)$ is 
\begin{equation}
\label{gamma}
\g:g\l(x\r) \mapsto g\l(\g\l(x\r)\r)= e^{i\alpha_1Q_{12}}e^{i\alpha_2Q_{34}}e^{i\alpha_3Q_{56}}\cdot x=\g\cdot g\l(x\r),
\end{equation}
where the composition `$\cdot$' is the usual matrix multiplication.\footnote{This gives the embedding of $\G$ in $SO\l(6\r)$.} This section  has the property (\ref{compatible}) that we demand from a thermal section. Hence, we choose the thermal section to be 
\begin{equation}
\label{thermal}
\s_{th}\l(x\r)=g\l(x\r).
\end{equation}
This analysis can be repeated for any odd-dimensional sphere to find the same expression for the thermal section. Essentially the only difference is that for a $(2n+1)$-dimensional sphere, we need to consider $(n+1)$ complex numbers $z_i$ and proceed exactly as above. We find that the thermal section can always be chosen to be (\ref{thermal}).

\subsection{The Method of Images}

Since the heat kernel obeys a linear differential equation- the heat equation- we can use the method of images to construct the heat kernel on $\G\backslash G/H$ from that on $G/H$ (see, for example \cite{Giombi:2008vd}). The relation between the two heat kernels is
\begin{equation}
\label{kernelonquotients}
K^{\l(S\r)}_{\G}\l(x,y;t\r)=\sum_{\g\in\G}K^{\l(S\r)}\l(x,\g\l(y\r);t\r),
\end{equation}
where $K^{\l(S\r)}_{\G}$ is the heat kernel between two points $x$ and $y$ on $\G\backslash G/H$, $K^{\l(S\r)}$ is the heat kernel on $G/H$ and the spin indices have been suppressed. 
We shall use this relation to determine the traced heat kernel on the thermal quotient of $S^{2n+1}$.

\subsection{The Traced Heat Kernel on thermal $S^{2n+1}$}

We use the formalism developed above to evaluate the traced heat kernel for the thermal quotient of the odd-dimensional sphere. Using the method of images, the quantity of interest is
\begin{equation}
K^{\l(S\r)}_{\G}\l(t\r)=\sum_{k\in\mathbb Z_N} \int_{\G\backslash G/H}d\mu\l(x\r)\sum_a K^{\l(S\r)}_{aa}\l(x,\g^k\l(x\r);t\r),
\end{equation}
where $d\mu(x)$ is the measure on $\G\backslash G/H$ obtained from the Haar measure on $G$, and $x$ labels points in $\G\backslash G/H$. 
We have also used the fact that $\G\simeq\mathbb Z_N$, and since the sum over $m$ is a finite sum, the integral has also been taken through the sum.
Further, since 
\begin{equation}
\sum_a K^{\l(S\r)}_{aa}\l(\g x,\g^k\l(\g x\r);t\r)= \sum_a K^{\l(S\r)}_{aa}\l(x,\g^k\l(x\r);t\r),
\end{equation}
the integral over $\G\backslash G/H$ can be traded in for the integral over $G/H$. We therefore multiply by an overall volume factor, and evaluate 
\begin{equation}
\int_{G/H} d\mu\l(x\r) \sum_a K^{\l(S\r)}_{aa}\l(x,\g^k\l(x\r);t\r),
\end{equation}
where $d\mu(x)$ is the left invariant measure on $G/H$ obtained from the Haar measure on $G$, and $x$ now labels points in the full coset space $G/H$. Putting the expression (\ref{kernel}) into this, and choosing the section (\ref{thermal}) we obtain
\begin{equation}
\int_{G/H} d\mu\l(x\r) \sum_a K^{\l(S\r)}_{aa}\l(x,\g^k\l(x\r);t\r)=\int_{G/H} d\mu\l(x\r)\sum_R  a_R^{(S)} \tr_S\l(g_x^{-1}\g^k g_x\r)^{\l(R\r)} e^{tE_R^{(S)}},
\end{equation} 
where $\l(g_x^{-1}\g^k g_x\r)^{\l(R\r)}$ is an abbreviation for ${\cu^{\l(R\r)}}\l(g\l(x\r)^{-1}\g^k g\l(x\r)\r)$. As this expression stands, the trace is only over some subspace $\mathcal V_S \subset \mathcal V_R$ so the cyclic property of the trace cannot be used to annihilate the $g_x^{-1}$ with the $g_x$. To proceed further, we move the integral into the summation to obtain
\begin{equation}
\int_{G/H} d\mu\l(x\r) \sum_a K^{\l(S\r)}_{aa}\l(x,\g^k\l(x\r);t\r)=\sum_R  a_R^{(S)} \int_{G/H} d\mu\l(x\r)\tr_S\l(g_x^{-1}\g^k g_x\r)^{\l(R\r)} e^{tE_R^{(S)}}.
\end{equation} 
Since $G$ and $H$ are compact, we may use the property that \cite{Helgason}
\begin{equation}
\label{integraldecomposition}
\int_G dg f\l(g\r)=\int_{G/H}d\mu(\tilde{x})\l[\int_H dh f\l(\tilde{x}h\r)\r],
\end{equation}
where $dg$ is the Haar measure on $G$, $d\mu$ and $dh$ are the invariant measures on $G/H$ and $H$ respectively. $\tilde{x}$ is an arbitrary choice of coset representatives that we make to label points in $G/H$. In what follows we shall choose the coset representative to be $g_x$.  Let us consider the function
\begin{equation}
f\l(g\r)= \tr_S\l(g^{-1}\g^k g\r)^{\l(R\r)}.
\end{equation}
This function has the property that $f\l(g_x h\r)=f\l(g_x\r)$.
Putting this in (\ref{integraldecomposition}), we see that the integral over $H$ becomes trivial, and we get
\begin{equation}
\label{integralrelation}
\int_{G/H} d\mu\l(x\r) \tr_S\l(g_x^{-1}\g^k g_x\r)^{\l(R\r)}=\frac{1}{V_H}\int_G dg \tr_S\l(g^{-1}\g^k g\r)^{\l(R\r)}.
\end{equation}
Now, as in \cite{David:2009xg}, we note that the integral $I_G=\int_G dg \l(g^{-1}\g^k g\r)$ commutes with all group elements $\tilde{g}\in G$ because
\begin{equation}
I_G\cdot \tilde{g}=\int_G dg \l(g^{-1}\g^k g\r)\cdot \tilde{g}=\tilde{g}\cdot \int_G d\l(g\tilde{g}\r) \l(\l(g\tilde{g}\r)^{-1}\g^k \l(g\tilde{g}\r)\r)=\tilde{g}\cdot I_G,
\end{equation}
where we have used the right invariance of the measure, viz. $dg=d\l(g\tilde{g}\r)$. We therefore have, from Schur's lemma, that
\begin{equation}
I_G=\int_G dg \l(g^{-1}\g^k g\r)\varpropto \mathbb{I},
\end{equation}
from which we obtain that
\begin{equation}
\label{tracerel}
\int_G dg \tr_S\l(g^{-1}\g^k g\r)^{\l(R\r)}=\frac{d_S}{d_R}\int_G dg \tr_{R}\l(g^{-1}\g^k g\r)^{\l(R\r)}
=\frac{d_S}{d_R}V_G \tr_R\l(\g^k\r).
\end{equation}
The quotient $\g$ is just the exponential of the Cartan generators of $SO\l(2n+2\r)$ (see, for example (\ref{gamma}) for the $S^5$). This trace, therefore is just the $SO\l(2n+2\r)$ character $\chi_R$ in the representation $R$. 
Putting this result in (\ref{integralrelation}), we find that
\begin{equation}
\int_{G/H} d\mu\l(x\r) \tr_S\l(g_x^{-1}\g^k g_x\r)^{\l(R\r)}=V_{G/H}\frac{d_S}{d_R}\chi_{R}\l(\g^k\r),
\end{equation}
where we have normalised volumes so that $V_{G}=V_{G/H}V_{H}.$ Therefore,
\begin{equation}
\sum_{k\in \bz_N}\int_{G/H} d\mu\l(x\r)K^{\l(S\r)}\l(x,\g^k\l(x\r);t\r)
=\sum_{k\in \bz_N}\sum_R \chi_{R} \l(\g^k\r) e^{tE_R^{(S)}},
\end{equation} 
where we have inserted the value of the normalisation constant $a_R^{\l(S\r)}$ from (\ref{normalisation}). Now for the thermal quotient, we have $\g$ such that
\begin{equation}
\label{thermalsphere}
\alpha_1\neq 0,\, \alpha_i=0,\quad\forall\, i=2,\ldots,n+1.
\end{equation}
The volume factor for this quotient is just $\frac{\alpha_1}{2\pi}$. This gives us the traced heat kernel on the thermal $S^{2n+1}$
\begin{equation}
\label{s5intkernel}
K^{\l(S\r)}_{\G}\l(t\r) =\frac{\alpha_1}{2\pi}\sum_{k\in \bz_N}\sum_R \chi_{R} \l(\g^k\r) e^{tE_R^{(S)}}.
\end{equation}
We find that the answer assembles naturally into a sum of characters, in various representations of $G$, of elements of the quotient group $\G$. Here the representations $R$ of $SO(2n+2)$ are those which contain $S$ when restricted to $SO(2n+1)$. The reader should compare this expression to the equation $\l(4.20\r)$ obtained in \cite{David:2009xg}.

\section{The Heat Kernel on $AdS_{2n+1}$}

\label{kernelAdS}

We have so far obtained an expression for the heat kernel on a compact symmetric space (\ref{untracedkernel}) and have extended our analysis to its left quotients. In particular, we have shown how the traced heat kernel on (a class of) quotients of $S^{2n+1}$ assembles into a sum over characters of the orbifold group $\G$. We now extend our analysis to the hyperbolic space ${\mathbb{H}}_{2n+1}$. Following the analysis of \cite{David:2009xg}, we will use the fact that the $N$-dimensional sphere admits an analytic continuation to the corresponding euclidean $AdS$ geometry. We now give an account of how one can exploit this fact to determine the heat kernel on $AdS_{2n+1}$ and its quotients.

\subsection{Preliminaries}

Euclidean $AdS$ is the $N$ dimensional hyperbolic space ${\mathbb H}_N^+$ which is the coset space
\begin{equation}
{\mathbb H}_N\simeq SO\l(N,1\r)/SO\l(N\r),
\end{equation} 
where the quotienting is done, as in the sphere, by the right action. As was done for the three-dimensional case, we will view $SO\l(N,1\r)$ as an analytic continuation of  $SO\l(N+1\r)$. 

For explicit expressions, we shall employ the generalisation of the triple-polar coordinates that we introduced in (\ref{s5pt}) to the general $S^{2n+1}$. In these coordinates, the $S^{2n+1}$ metric is
\begin{equation}
\label{s5metric}
d\theta^2+cos^2\theta d\phi_1^2 + sin^2\theta\, d\Omega^2_{2n-1}.
\end{equation}
We perform the analytic continuation 
\begin{equation}
\label{continuation}
\theta \mapsto -i\rho, \quad \phi_1 \mapsto it,
\end{equation} where $\rho$ and $t$ take values in $\mathbb{R}$, to obtain
\begin{equation}
ds^2=-\l(d\rho^2+cosh^2\rho dt^2 + sinh^2\rho\, d\Omega^2_{2n-1}\r).
\end{equation}
The reader will recognize this as the metric on global $AdS_{2n+1}$, upto a sign. 

Now, to construct eigenfunctions on ${\mathbb{H}}_N$, we need to write down a section in $SO(N,1)$. The Lie algebra of $SO(N,1)$ is an analytic continuation of $SO(N+1)$ where we choose a particular axis- say `1'- as the time direction and perform the analytic continuation $Q_{1j}\rightarrow iQ_{1j}$ to obtain the $so(N,1)$ algebra from the $so(N+1)$ algebra. This is equivalent to the analytic continuation of the coordinates described above. Therefore, the section in $SO(N,1)$ can be obtained from that in $SO\l(N+1\r)$ by analytically continuing the coordinates via (\ref{continuation}).

\subsection{Harmonic Analysis on ${\mathbb{H}}_{2n+1}$}
\label{HarmonicAnalysisAdS}

We have recollected basic results from harmonic analysis on coset spaces in Section \ref{HarmonicAnalysis}, which we have exploited for compact groups $G$ and $H$. In fact, all the basic ingredients that we have employed in our analysis can be carried over to the case of non-compact groups as well. The eigenvalues of the spin-$S$ Laplacian are still given by (\ref{eigenval}), and the eigenfunctions are still (\ref{eigenf}), \textit{i.e.} they are determined by matrix elements of unitary representations of $G$. These unitary representations are now infinite dimensional, given that $G$ is non-compact. However, for $SO(N,1)$, these representations have been classified \cite{Hirai,Camporesi}, and we shall use these results to determine the traced heat kernel on $AdS_{2n+1}$.

The only unitary representations of $SO(N,1)$ that are relevant to us are those that contain unitary representations of $SO(N)$. For odd-dimensional hyperboloids, where $N=2n+1$, these are just the so-called principal series representations of $SO\l(2n+1,1\r)$ which are labelled by the array 
\begin{equation}
\label{principalseries}
R=\l(i\lambda,m_2,m_3,\cdots,m_{n+1}\r), \quad \lambda \in {\mathbb R},\,m_2\geq m_3\geq\cdots\geq m_n\geq\vert m_{n+1}\vert, 
\end{equation} where the $m_2,\cdots ,m_n$ and $\vert m_{n+1}\vert$ are non-negative (half-)integers. We shall usually denote the array $\l(m_2,m_3,\ldots,m_{n+1}\r)$ by $\vec{m}$. We also note that the principal series representations $\l(i\lambda,\vec{m}\r)$ that contain a representation $S$ of $SO\l(2n+1\r)$ are determined by the branching rules \cite{Ottoson,Camporesi}
\begin{equation}
\label{adsbranching}
s_1 \geq m_2\geq s_2\geq \ldots \geq m_n\geq s_n\geq\vert m_{n+1}\vert,
\end{equation}
which, for the STT tensors just reduce to $m_2=s$ with all other $m_i$s and $s_i$s set to zero (\textit{cf.} \ref{snbranchingSTT}), except for the case of $n=1$, where the branching rule is $\vert m_2\vert=s$.
Comparing (\ref{principalseries}) to (\ref{so2nrep}) suggests that the appropriate analytic continuation is 
\begin{equation}
m_1\mapsto i\lambda-n, \quad \lambda\in\mathbb{R}_+,
\end{equation}
which is indeed the analytic continuation used by \cite{Camporesi:1994ga}.
Let us consider how the eigenvalues $E_R^{\l(S\r)}$ transform under this analytic continuation. It turns out that the eigenvalues (\ref{eigenvals2n1}) get continued to
\begin{equation}
E_{R,AdS_{2n+1}}^{\l(S\r)}=-\l(\gl^2+\zeta\r),\quad \zeta\equiv C_2\l(S\r)-C_2\l(\vec{m}\r)+n^2,
\end{equation}
and that the eigenvalue for the STT tensors (\ref{eigenvals2n1stt}) gets continued to
\begin{equation}
E_{R,AdS_{2n+1}}^{\l(S\r)}=-\l(\lambda^2+ s+n^2\r).
\end{equation}
The eigenvalues on $AdS$ have an extra minus sign apart from what is obtained by the analytic continuation because the metric $S^{2n+1}$ under the analytic continuation goes to minus of the metric on $AdS_{2n+1}$.
This analytic continuation preserves the corresponding energy eigenvalue as a negative definite real number, which it must, because the Laplacian on Euclidean $AdS$ is an elliptic operator, and its eigenvalues must be of definite sign.

\subsection{The coincident Heat Kernel on $AdS_{2n+1}$}

In computing the heat kernel over $AdS_{2n+1}$ by analytically continuing from $S^{2n+1}$, the sum over $m_1$ that entered in (\ref{kernelsphere}), (\ref{kernelsttsphere}) and (\ref{kerneltracesttsphere}) is now continued to an integral over $\lambda$. In general this integral over $\lambda$ is hard to perform, but it simplifies significantly in the coincident limit and is evaluated below for this case. The traced heat kernel for STT tensors has previously been obtained directly in this limit by \cite{Camporesi:1994ga} and this calculation therefore serves as a check of the prescription (\ref{continuation}) of analytic continuation that we have employed. We will also see that the normalisation constant $a_{R}^S$ that appeared for the $S^{2n+1}$ gets continued to $\mu_{R}^S$, which is essentially the measure for this integral. This is a brief summary of the calculation, the reader will find more details in Appendix \ref{plancherelcalc}.

On using (\ref{dimensions}) for the special case of $R=\l(m,s,0,\ldots,0\r)$, one can show that $a_R^S$ gets continued via (\ref{continuation}) to $\mu_R^S$, where
\begin{equation}
\mu_R^S=\frac{1}{d_s}\frac{\l[\gl^2+\l(s+\frac{N-3}{2}\r)^2\r]\prod_{j=0}^{\frac{N-5}{2}}\l(\gl^2+j^2\r)} {2^{N-1}\pi^{\frac{N}{2}}\G\l(\frac{N}{2}\r)} \frac{\l(2s+N-3\r)\l(s+N-4\r)!}{s!\l(N-3\r)!},
\end{equation}
where $N=2n+1$. A little algebra reveals this as the combination
\begin{equation}
\mu_R^S=\frac{C_N g\l(s\r)}{d_S}\frac{\mu\l(\gl\r)}{\Omega_{N-1}},
\end{equation}
in the notation of \cite{Camporesi:1994ga}, (see their expressions $2.7$ to $2.13$ and $2.108$ or our Appendix \ref{plancherelcalc}). We have omitted the overall sign $\l(-1\r)^{\frac{N-1}{2}}$ in writing the above. The quantity $\mu\l(\gl\r)$ is known as the Plancherel measure.
We now consider the expression (\ref{kerneltracesttsphere}) on $S^{2n+1}$, which in the coincident limit, reduces to
\begin{equation}
K^{S}\l(x,x;t\r) = \sum_{m_1} a_R^S d_S e^{tE_R^{(S)}}.
\end{equation}
where $a_R^S d_S=\frac{n!}{2\pi^{n+1}}d_{\l(m_1,s\r)}$. We analytically continue this expression via our prescription (\ref{continuation}) to the coincident heat kernel on $AdS_{2n+1}$. 
\begin{equation}
K^{S}\l(x,x;t\r) = \int d\gl\, \mu_R^S\l(\gl\r) d_S e^{tE_R^{(S)}}= \frac{C_N}{\Omega_{N-1}} g\l(s\r) \int d\gl\,\mu\l(\gl\r) e^{-t\l(\lambda^2+ s+n^2\r),}
\end{equation}
which is precisely the expression obtained by \cite{Camporesi:1994ga}.

\section{The Heat Kernel on Thermal $AdS_{2n+1}$}
\label{kernelthermalads}
\subsection{The Thermal Quotient of $AdS$}

We are now in a position to calculate the traced heat kernel of an arbitrary tensor particle on thermal $AdS_{2n+1}$. This space is the hyperbolic space ${\mathbb H}_{2n+1}$ with a specific $\mathbb Z$ identification (in the generalised polar coordinates)
\begin{equation}
t\sim t+\beta,\quad \beta=i\alpha_1
\end{equation}
which is just the analytic continuation by (\ref{continuation}) of the identification (\ref{thermalsphere}) on the sphere. Since $t$ is the global time coordinate, $\beta$ is to be interpreted as the inverse temperature. 

\subsection{The Heat Kernel}

In section \ref{kernelAdS} we have discussed how the heat kernel on ${\mathbb H}_{2n+1}$ can be calculated by analytically continuing the harmonic analysis on $S^{2n+1}$ to ${\mathbb H}_{2n+1}$. As discussed in Sec. 6.2 of \cite{David:2009xg}, we expect to be able to continue the expressions for the heat kernel on the thermal sphere to thermal $AdS$, with the difference that now $\G\simeq\mathbb Z$, rather than $\mathbb Z_N$. Also, as noted in \cite{David:2009xg}, essentially the only difference that arises for the traced heat kernel is that the character of $SO(2n+2)$ that appears in (\ref{s5intkernel}) is now replaced by the Harish-Chandra (or global) character for the non-compact group $SO(2n+1,1)$. 

With these inputs, the traced heat kernel on thermal $AdS_{2n+1}$ is given by
\begin{equation}
\label{ads5intkernel}
K^{\l(S\r)}\l(\gamma,t\r)
=\frac{\beta}{2\pi}\sum_{k\in \bz}\sum_{\vec{m}}\int_0^\infty d\lambda\, \chi_{\lambda,\vec{m}} \l(\g^k\r) e^{tE_R^{(S)}},
\end{equation}
where $\chi_{\lambda,\vec{m}}$ is the Harish-Chandra character in the principal series of $SO\l(2n+1,1\r)$, which has been evaluated \cite{HiraiChar} to be
\begin{equation}
\label{fullHCchar}
\chi_{\lambda,\vec{m}}\l(\beta,\phi_1,\phi_2,\ldots,\phi_n\r)=
\frac{e^{-i\beta\lambda}\chi^{SO\l(2n\r)}_{\vec{m}}\l(\phi_1,\phi_2,\ldots,\phi_n\r)+e^{i\beta\lambda}\chi^{SO\l(2n\r)}_{\check{\vec{m}}}\l(\phi_1,\phi_2,\ldots,\phi_n\r)}{e^{-n\beta}\prod_{i=1}^n\vert e^\beta - e^{i\phi_i}\vert^2},
\end{equation}
for the group element
\begin{equation}
\g=e^{i\beta Q_{12}}e^{i\phi_1 Q_{23}}\ldots e^{i\phi_n Q_{2n+1,2n+2}},
\end{equation}
where ${\check{\vec{m}}}$ is the conjugated representation, with the highest weight $\l(m_2,\ldots,-m_{n+1}\r)$, and $\chi^{SO\l(2n\r)}_{\vec{m}}$ is the character in the representation $\vec{m}$ of $SO\l(2n\r)$.
The sum over $\vec{m}$ that appears in (\ref{ads5intkernel}) is the sum over permissible values of $m$ as determined by the branching rules (\ref{adsbranching}). We also recall that `1' is the time-like direction.

For the thermal quotient, we have $\beta\neq 0$ and $\phi_i=0\;\forall\, i$. The character of $SO\l(2n\r)$ that appears in the character formula (\ref{fullHCchar}) above is then just the dimension of the corresponding representation. Using the fact that the dimensions of this representation is equal to that of its conjugate, we have for the character
\begin{equation}
\label{thermalHCchar}
\chi_{\lambda,\vec{m}}\l(\beta,\phi_1,\phi_2,\ldots,\phi_n\r)=
\frac{cos\l(\beta\lambda\r)}{2^{2n-1}\sinh^{2n}{\beta\over 2}}d_{\vec{m}}.
\end{equation}
We therefore obtain, for the traced heat kernel $AdS_{2n+1}$ (\ref{ads5intkernel}), 
\begin{equation}
K^{\l(S\r)}\l(\beta,t\r)
=\frac{\beta}{2^{2n}\pi}\sum_{k\in \bz}\sum_{\vec{m}}d_{\vec{m}}\int_0^\infty d\lambda\, \frac{cos\l(k\beta\lambda\r)}{\sinh^{2n}{k\beta\over 2}} e^{-t\l(\lambda^2+\zeta\r)}.
\end{equation}
The integral over $\lambda$ is a Gaussian integral, which we can evaluate to obtain
\begin{equation}
K^{\l(S\r)}\l(\beta,t\r)
=\frac{\beta}{2^{2n}\sqrt{\pi t}}\sum_{k\in \bz_+}\sum_{\vec{m}}d_{\vec{m}}\frac{1}{\sinh^{2n}{k\beta\over 2}}e^{-\frac{k^2 \beta ^2}{4t}-t\zeta},
\end{equation}
where we have dropped the term with $k=0$, which diverges. This divergence arises due to the infinite volume of $AdS$, over which the coincident heat kernel on the full $AdS_{2n+1}$ is integrated. It can be reabsorbed into a redefinition of parameters of the gravity theory under study and is independent of $\beta$ and is therefore not of interest to us. 

This expression further simplifies for the case of the STT tensors. The branching rules determine that $\vec{m}=\l(s,0,\ldots,0\r)$ and therefore the sum over $\vec{m}$ gets frozen out and we obtain
\begin{equation}
K^{\l(S\r)}\l(\beta,t\r)
=\frac{\beta}{2^{2n}\sqrt{\pi t}}\sum_{k\in \bz_+}\frac{d_{\vec{m}}}{\sinh^{2n}{k\beta\over 2}}e^{-\frac{k^2 \beta ^2}{4t}-t\l(s+n^2\r)}.
\end{equation}
The reader may compare this expression to the equation $\l(3.9\r)$ obtained in \cite{David:2009xg} for the $AdS_3$ case (where one would have to specialise to $\t_1=0$, $\t_2=\beta$, and $d_{\vec{m}}\equiv 1$, and further include a factor of $2$ that appears because the branching rule for $AdS_3$ leads us to sum over $m_2=\pm s$ rather than $m_2=s$ for $s>0$).
\section{The one-loop Partition Function}

As a consequence of the above, we can calculate the one-loop determinant of a spin-$S$ particle on $AdS_5$. To do so, we need the result that
\begin{equation}
\int_0^\infty \frac{dt}{t^{\frac{3}{2}}}e^{-\frac{\alpha^2}{4t}-\beta^2t}=\frac{2\sqrt{\pi}}{\alpha}e^{-\alpha\beta}.
\end{equation}
Then the one-loop determinants can be deduced from the heat kernel by using
\begin{equation}
-log\,det\l(-\Delta_{\l(S\r)}+m_S^2\r)=\int_0^\infty \frac{dt}{t}K^{\l(S\r)}\l(\beta,t\r)e^{-m_S^2t},
\end{equation}
which we can simplify to obtain
\begin{equation}
-log\,det\l(-\Delta_{\l(S\r)}+m_S^2\r)=\sum_{k\in \bz_+}\sum_{\vec{m}}d_{\vec{m}}\frac{2}{e^{-nk\beta}\l( e^{k\beta} -1\r)^{2n}}\frac{1}{k}e^{-k\beta\sqrt{\zeta+m_S^2}}.
\end{equation}
This expression further simplifies for the case of STT tensors and one obtains
\begin{equation}
-log\,det\l(-\Delta_{\l(S\r)}+m_S^2\r)=\sum_{k\in \bz_+}d_{\vec{m}}\frac{2}{e^{-nk\beta}\l( e^{k\beta} -1\r)^{2n}}\frac{1}{k}e^{-k\beta\sqrt{s+n^2+m_S^2}}.
\end{equation}
\subsection{The scalar on $AdS_5$}

Let us evaluate the above expression for scalars in $AdS_5$, where $s=0$. In units where the $AdS$ radius is set to one, $\sqrt{m_S^2+4}=\Delta-2$, where $\Delta$ is the conformal dimension of the scalar. We therefore have
\begin{equation}
-log\,det\l(-\Delta_{\l(S\r)}+m_S^2\r)=\sum_{k\in \bz_+}\frac{2}{k\l(1-e^{-k\beta}\r)^{4}}e^{-k\beta\D}
\end{equation}
We can evaluate the sum to find that the one-loop determinant is given by
\begin{equation}
-log\,det\l(-\Delta_{\l(S\r)}+m_S^2\r)=-2\sum_{n=0}^\infty \frac{\l(n+1\r)\l(n+2\r)\l(n+3\r)}{6}log\l(1-e^{-\beta\l(\Delta+n\r)}\r).
\end{equation}
Now since $logZ_{\l(S\r)}=-\frac{1}{2}log\,det\l(-\Delta_{\l(S\r)}+m_S^2\r)$, we have, for the one-loop partition function of a scalar,
\begin{equation}
\label{scalarpartition}
logZ_{\l(S\r)}=-\sum_{n=0}^\infty \frac{\l(n+1\r)\l(n+2\r)\l(n+3\r)}{6}log\l(1-e^{-\beta\l(\Delta+n\r)}\r).
\end{equation}
This expression matches exactly with that which is obtained with the method of \cite{Denef:2009kn} (see also, earlier work by \cite{Hartman:2006dy,Diaz:2007an}).

\textit{Note:} After version 1 of this paper appeared on the arXiv, we learned of \cite{Gibbons:2006ij} where expressions for the one-loop partition function for general spin on thermal $AdS$ were obtained by means of a Hamiltonian computation\footnote{We thank Gary Gibbons for bringing this to our attention.}. Our results agree with the expressions obtained there.

\section{Conclusions}

We have computed the principal ingredients that go into the calculation of one loop effects on odd dimensional thermal AdS spacetimes. As mentioned in the introduction, there are many potential applications of these results. Specifically, in the context of investigating higher spin gauge fields in these spacetimes. It would also be useful to complete the analysis for the even dimensional AdS spacetimes as well. Finally, the ambitious goal, which was the initial motivation for this work, is to obtain some clues about the one loop partition function of the string theory on AdS. The heat kernel method is ideally suited for this purpose and we expect our explicit results will be helpful in this regard.

\section*{Acknowledgments}

We would like to thank M. Gaberdiel for helpful comments on the manuscript. We would also like to thank Shamik Banerjee, Andrea Campoleoni, Justin David, Dileep Jatkar, and Suvrat Raju for helpful discussions. The work of R.G. was supported by a Swarnajayanthi Fellowship of the DST, Govt. of India and more generally by the generosity of the Indian people. The work of R. Gupta is a part of research programme of FOM, which is financially supported by the Netherlands Organization for Scientific Research (NWO).
\appendix
\section*{Appendix}
\section{Normalising the Heat Kernel on $G/H$}
\label{normalisationkernel}
We determine the normalisation factor $a^{\l(S\r)}_R$ that arises in the expressions (\ref{kernelintermsofeigenfunctions}) onwards. This is fixed by demanding that the integrated traced heat kernel obey
\begin{equation}
\label{normalisation1}
\int_{G/H}d\mu\l(x\r)K^{(S)}\l(x,x;t\r)=\sum_R d_R e^{tE_R^{(S)}}.
\end{equation}
Using the expression (\ref{untracedkernel}) for the heat kernel on $G/H$, we have
\begin{eqnarray}
\label{normalisation2}
\int_{G/H}d\mu\l(x\r)K^{(S)}\l(x,x;t\r)&=&\sum_R \int_{G/H}d\mu\l(x\r)  a_R^{(S)} {{\cu^{\l(R\r)}}\l(\s\l(x\r)^{-1}\s\l(x\r)\r)_a}^a e^{tE_R^{(S)}}\nonumber\\&=& \sum_R a_R^{(S)} V_{G/H}  d_S e^{tE_R^{(S)}}.
\end{eqnarray}
On comparing (\ref{normalisation1}) and (\ref{normalisation2}), we obtain the required relation
\begin{equation}
\label{normalisation}
a_R^{(S)}=\frac{d_R}{V_{G/H} d_S},
\end{equation} 
which we use in the main text from (\ref{kernelintermsofeigenfunctions}) onwards.

\section{The Plancherel Measure for STT tensors on $H_N$}
\label{plancherelcalc}
We show how the normalisation constant $a_R^S$ gets analytically continued to $\mu_R^S$, the measure for the $\gl$ integration that appears in the $AdS$ heat kernel. Let us consider the expression for $d_{\l(m,s\r)}$ which we obtain from (\ref{dimensions}).
\begin{equation}
d_{m_1,s}=\prod_{j=2}^{n+1}\frac{l_1^2-l_j^2}{\mu_1^2-\mu_j^2}\prod_{j=3}^{n+1}\frac{l_2^2-l_j^2}{\mu_2^2-\mu_j^2}.
\end{equation}
The first product in the numerator gets analytically continued via (\ref{continuation}) to
\begin{equation}
\prod_{j=2}^{n+1}\l(l_1^2-l_j^2\r)\mapsto \l(-1\r)^{n} \l[\gl^2+\l(s+n-1\r)^2\r]\prod_{j=0}^{n-2}\l(\gl^2+j^2\r),
\end{equation}
while the second product evaluates to 
\begin{equation}
\prod_{j=2}^{n+1}\l(l_2^2-l_j^2\r)=\frac{\l(s+n-1\r)\l(s+2n-3\r)!}{s!}.
\end{equation}
The denominator $\prod_{j=2}^{n+1}\l(\mu_1^2-\mu_j^2\r)\prod_{j=3}^{n+1}\l(\mu_1^2-\mu_j^2\r)$ evaluates to
\begin{equation}
\prod_{j=2}^{n+1}\l(\mu_1^2-\mu_j^2\r)\prod_{j=3}^{n+1}\l(\mu_1^2-\mu_j^2\r)=\l(2n-2\r)!\frac{2^{2n-2}}{\sqrt{\pi}} n!\G\l(n+\frac{1}{2}\r).
\end{equation}
The dimension $d_{\l(m,s\r)}$ then gets continued to
\begin{equation}
\l(-1\r)^{\frac{N-1}{2}} \frac{\l[\gl^2+\l(s+\frac{N-3}{2}\r)^2\r]\prod_{j=0}^{\frac{N-5}{2}}\l(\gl^2+j^2\r)} {\frac{2^{N-2}}{\sqrt{\pi}}\G\l(\frac{N}{2}\r)\l(\frac{N-1}{2}\r)!} \frac{\l(2s+N-3\r)\l(s+N-4\r)!}{s!\l(N-3\r)!},
\end{equation}
where we have changed variables from $n$ to $N=2n+1$. We now use the fact that for odd $N$
\begin{equation}
V_{S^N}=\frac{\l(N+1\r)\pi^{\frac{N+1}{2}}}{\G\l(\frac{N+3}{2}\r)}=\frac{2\pi^{\frac{N+1}{2}}}{\G\l(\frac{N+1}{2}\r)}, =\frac{2\pi^{\frac{N+1}{2}}}{\l(\frac{N-1}{2}\r)!}
\end{equation}
and hence the combination $\frac{d_{\l(m,s\r)}}{V_{G/H}}$ gets mapped to
\begin{equation}
\l(-1\r)^{\frac{N-1}{2}} \frac{\l[\gl^2+\l(s+\frac{N-3}{2}\r)^2\r]\prod_{j=0}^{\frac{N-5}{2}}\l(\gl^2+j^2\r)} {2^{N-1}\pi^{\frac{N}{2}}\G\l(\frac{N}{2}\r)} \frac{\l(2s+N-3\r)\l(s+N-4\r)!}{s!\l(N-3\r)!}.
\end{equation}
Using the expressions from \cite{Camporesi:1994ga} quoted in the main text, \textit{i.e.}
\begin{equation}
\Omega_{N-1}=\frac{2\pi^{\frac{N}{2}}}{\Gamma\l(\frac{N}{2}\r)},\quad c_N=\frac{2^{N-2}}{\pi},\quad g(s)=\frac{(2s+N-3)(s+N-4)!}{(N-3)!s!}
\end{equation}
and
\begin{equation}
\mu(\lambda)=\frac{\pi[\lambda^2+(s+\frac{N-3}{2})^2]\prod_{j=0}^{\frac{N-5}{2}}(\lambda^2+j^2)}{\l[2^{N-2}\G\l(\frac{N}{2}\r)\r]^2},
\end{equation}
we see that the normalisation constant $a_R^S$ gets mapped to 
\begin{equation}
\mu_R^S=\frac{C_N g\l(s\r)}{d_S}\frac{\mu\l(\gl\r)}{\Omega_{N-1}},
\end{equation}
where we have omitted the overall sign that appears for some values of $N$ as an artefact of the analytic continuation, since the measure is always positive definite.
The coincident heat kernel is therefore
\begin{equation}
K^{S}\l(x,x,t\r) = \int d\gl\, \mu_R^S\l(\gl\r) d_S e^{tE_R^{(S)}}= \frac{C_N}{\Omega_{N-1}} g\l(s\r) \int d\gl\,\mu\l(\gl\r) e^{-t\l(\lambda^2+ s+n^2\r)}.
\end{equation}
Now, the coincident heat kernel may also be written down using $2.7$ of \cite{Camporesi:1994ga} (in their notation) as
\begin{equation}
K^{S}\l(x,x,t\r)=\sum_u \int_0^\infty d\gl\, \hat{h}^{\gl u*}\cdot\hat{h}^{\gl u}\l(x\r)e^{-t\l(\lambda^2+ s+n^2\r)}.
\end{equation}
On choosing $x$ to be the origin (which we can do for arbitrary $x$), and using their expression $2.10$, we conclude that
\begin{equation}
K^{S}\l(x,x,t\r) =\frac{C_N}{\Omega_{N-1}} g\l(s\r) \int d\gl\,\mu\l(\gl\r) e^{-t\l(\lambda^2+ s+n^2\r)},
\end{equation}
which is precisely the expression we have obtained via analytic continuation.
\section{The Dictionary for the $S^3$ Calculation}
Let us consider the expressions derived for the $S^3$ in the $\l(SU\l(2\r) \times SU\l(2\r)\r)/SU\l(2\r)$ language adopted in \cite{David:2009xg}. We would like to show that they coincide with our expressions, when recast in the $SO\l(4\r)/SO\l(3\r)$ language. In particular we shall show that the choice for the thermal section made previously for $S^3$ is indeed $\s_{th}\l(x\r)=g\l(x\r)$ as has been made here.

The Lie algebra of $SU\l(2\r) \times SU\l(2\r)$ is spanned by $\l(\cj_i,\ck_i\r)$, that are linear combinations of the $Q_{\l(ij\r)}$'s that span the lie algebra of $SO\l(4\r)$.
\begin{eqnarray}
\cj^z = \frac{1}{2}\l(Q_{12}+Q_{34}\r) &\quad& 
\ck^z = \frac{1}{2}\l(Q_{12}-Q_{34}\r) \nonumber \\
\cj^+ = \frac{1}{2}\l(Q_{24}+Q_{31}+i\l(Q_{32}-Q_{14}\r)\r) &\quad&
\ck^+ = \frac{1}{2}\l(Q_{24}-Q_{31}-i\l(Q_{14}+Q_{32}\r)\r) \nonumber \\
\cj^- = \cj^{+\dagger} &\quad& \ck^- = \ck^{+\dagger}. 
\end{eqnarray}
We note that the relations between the two Cartans $\cj^z$ and $\ck^z$ give us the following dictionary between highest weights written in the $SO\l(4\r)$ language and in the $SU\l(2\r) \times SU\l(2\r)$ language
\begin{equation}
\label{dictionaryS3}
\l(j_1,j_2\r)_{SU\l(2\r) \times SU\l(2\r)}\equiv \l(j_1+j_2,j_1-j_2\r)_{SO\l(4\r)}.
\end{equation}
Let us now consider the coset representative in double polar coordinates in the $SU\l(2\r)$ representation. It is given by (see equation $\l(2.7\r)$ of \cite{David:2009xg})
\begin{equation}
x\equiv g\l(\psi,\eta,\phi\r)=\l(\begin{array}{cc} e^{-i\eta}\cos\psi & ie^{i\phi}\sin\psi \\ ie^{-i\phi}\sin\psi & e^{i\eta}\cos\psi\\ \end{array} \r)
\end{equation}
To translate the expressions of \cite{David:2009xg} to our form, it is useful to embed the $S^3$ into ${\mathbb R}^4$ via the coordinates $\l(x_1,x_2,x_3,x_4\r)$. The identification is through
\begin{equation}
x =\l(\begin{array}{cc} e^{-i\eta}\cos\psi & ie^{i\phi}\sin\psi \\ ie^{-i\phi}\sin\psi & e^{i\eta}\cos\psi\\ \end{array} \r)\equiv\l(\begin{array}{cc} x_1-ix_2 & ix_3-x_4 \\ ix_3+x_4 & x_1+ix_2\\ \end{array} \r),
\end{equation}
from which we read off the coordinates
\begin{eqnarray}
\label{s3pt}
x_1=\cos\eta \, \cos\psi &\quad& x_2=\cos\eta \, \sin\psi, \nonumber \\
x_3=\cos\phi \, \sin\psi &\quad& x_4=\sin\phi \, \sin\psi.
\end{eqnarray}
With this identification, the north pole $x_\circ$ of $S^3$ is given by the matrix
\begin{equation}
x_\circ=\mathbb{I}.
\end{equation}
The thermal section in \cite{David:2009xg} was chosen to be the pair of matrices $\l(g_L\l(x\r),g_R\l(x\r)\r)$ given by (see equations $\l(2.30\r)$ and $\l(2.31\r)$ of \cite{David:2009xg}),
\begin{equation}
g_L\l(\psi,\eta,\phi\r)=\l(\begin{array}{cc} e^{i\l(\phi-\eta\r)/2}\cos\frac{\psi}{2} & ie^{i\l(\phi-\eta\r)/2}\sin\frac{\psi}{2} \\ ie^{-i\l(\phi-\eta\r)/2}\sin\frac{\psi}{2} & e^{-i\l(\phi-\eta\r)/2}\cos\frac{\psi}{2}\\ \end{array} \r),
\end{equation}
and,
\begin{equation}
g_R\l(\psi,\eta,\phi\r)=\l(\begin{array}{cc} e^{i\l(\phi+\eta\r)/2}\cos\frac{\psi}{2} & -ie^{i\l(\phi+\eta\r)/2}\sin\frac{\psi}{2} \\ -ie^{-i\l(\phi+\eta\r)/2}\sin\frac{\psi}{2} & e^{-i\l(\phi+\eta\r)/2}\cos\frac{\psi}{2}\\ \end{array} \r).
\end{equation}
An element $\l(A,B\r)$ of $G$ acts on cosets $x\in G/H$ via
\begin{equation}
x\mapsto AxB^{-1},
\end{equation}
and therefore the thermal section maps the north pole to the point
\begin{equation}
x_{\circ}\mapsto g_L\l(x\r)x_{\circ}g_{R}^{-1}\l(x\r)=g_L\l(x\r)g_{R}^{-1}\l(x\r)=x.
\end{equation}
Therefore, in the language of $SO\l(4\r)/SO\l(3\r)$, the thermal section is the matrix that rotates the north pole to the point (\ref{s3pt}). This is precisely the choice (\ref{thermal}) we would have made for the thermal section using our geometric construction. 
\section{An extension to Even Dimensions}
\label{evendim}
We have so far considered the case of the odd-dimensional hyperboloids. This is mainly because we have obtained the heat kernel answer by means of an analytic continuation from the sphere which essentially captures the `principal series' contribution to the heat kernel. For the odd-dimensional case there is no other contribution. The case of even-dimensional hyperboloids is however a bit more subtle. There can in principle be a contribution from the discrete series also. However, as we outline below, this series does not contribute for a wide class of tensor fields. In particular, this includes the case of the STT tensors which have been of special interest to us\footnote{These remarks are true for higher (than two)-dimensional hyperboloids. There is an additional discrete series contribution in $AdS_2$ \textit{even} for the STT tensors. See \cite{Camporesi:1994ga} for details.}. The answer for the heat kernel in such cases is again captured by the usual analytic continuation, as was explicitly shown in \cite{Camporesi:1994ga}. 

We will now briefly sketch how the computation of the traced heat kernel would proceed for the even-dimensional hyperboloids. We begin by observing that the expression \refb{s5intkernel} is valid for cosets of compact Lie groups $G$ and $H$, which therefore includes the case of the even-dimensional spheres also. That a thermal section on such spheres may be defined is apparent via the geometric construction outlined in the main text\footnote{As an example, we see that setting $\phi_3=0$ in \refb{s5pt} gives us a parametrisation of $S^4$. The construction of the thermal section is then exactly analogous.}. The expression for the heat kernel on $S^{2n}\simeq SO\l(2n+1\r)/SO\l(2n\r)$, $n\geq 2$ is then a sum of characters of the `thermal' quotient group $\G$ embedded in $SO\l(2n+1\r)$.

The hyperboloid $AdS_{2n}$ is the quotient space $SO\l(2n,1\r)/SO\l(2n\r)$. The principal series of unitary irreducible representations of $SO\l(2n,1\r)$ are labelled by an array 
\begin{equation}
\label{principalserieseven}
R=\l(i\lambda,m_2,m_3,\cdots,m_{n}\r), \quad \lambda \in {\mathbb R},\,m_2\geq m_3\geq\cdots\geq m_n,
\end{equation} where the $m_2,\cdots ,m_n$ are non-negative (half-)integers. These contain a representation $S$ of $SO\l(2n\r)$ if \cite{Ottoson,Camporesi}
\begin{equation}
\label{adsbranchingeven}
s_1 \geq m_2\geq s_2\geq \ldots \geq m_n\geq \vert s_n\vert.
\end{equation}
The analytic continuation from the unitary irreducible representations of $SO\l(2n+1\r)$ to the principal series of $SO\l(2n,1\r)$ may be deduced as in Section \ref{HarmonicAnalysisAdS}. The answer is 
\begin{equation}
m_1\mapsto i\lambda-\frac{2n-1}{2}, \quad \lambda\in\mathbb{R}_+,
\end{equation}
which is precisely the continuation obtained in \cite{Camporesi:1994ga}.
There is in addition an additional discrete series of representations
which is not captured by this analytic continuation. However, this contains the representation $S$ only if $s_n\geq\frac{1}{2}$, see \cite{Camporesi} for details. Therefore, this additional series never contributes for the STT tensors (for which $s_n$ equals zero). The naive analytic continuation is therefore sufficient to give the full heat kernel answer.

The methods outlined in Sections \ref{kernelAdS} and \ref{kernelthermalads} may therefore be extended to even dimensional hyperboloids as well. The expressions for the global characters of $SO\l(2n,1\r)$ are well known \cite{HiraiChar}. We therefore have all the ingredients needed to compute the heat kernel on thermal $AdS_{2n}$.

For example, in this manner the one-loop partition function for a scalar on $AdS_4$ may be calculated. We find that
\begin{equation}
log Z_{(S)}=\sum_{k\in \bz_+}\frac{1}{k\l(1-e^{-k\beta}\r)^{3}}e^{-k\beta\D},
\end{equation}
where $\D$ is determined in terms of the mass of the scalar via 
\begin{equation}
\D=\sqrt{m^2+\frac{9}{4}}+\frac{3}{2}.
\end{equation}This matches, for instance, with the expressions obtained via the Hamiltonian analysis done in \cite{Gibbons:2006ij}.

\end{document}